\documentclass[prd,twocolumn,showpacs,floatfix]{revtex4}
\usepackage{graphicx}
\usepackage{amsfonts}
\usepackage{longtable}

\newcommand{\be}[0]{\begin{equation}}
\newcommand{\ee}[0]{\end{equation}}
\newcommand{\bea}[0]{\begin{eqnarray}}
\newcommand{\eea}[0]{\end{eqnarray}}

\begin{document}

\title{Decay channels and charmonium mass-shifts}

\author{M.R. Pennington and D.J. Wilson}

\affiliation{Institute for Particle Physics Phenomenology,\\ Durham University, 
Durham, DH1 3LE, U.K.}
\begin{abstract}
The discovery in the last few years of the $X, Y$ and $Z$ states of the 
extended charmonium family has highlighted the importance of the closeness
of decay channels to an understanding of these mesons.
We aid this debate by illustrating a simple calculational procedure
for including the effect of open and nearby closed channels.
\end{abstract}

\maketitle

\section{Modelling decay channels}

The discovery of narrow states of hidden charm, the $X,\,Y,\,Z$-mesons~\cite{swanson,pdg}, has generated a whole literature discussing their nature, structure and relation to charmonium. The fact that a state, like the $X(3872)$ sits between $D^{*0}{\overline{D^0}}$ and $D^{*+}D^-$ thresholds~\cite{footnote}, with a width of less than 1.2 MeV, has highlighted the potentially important role that hadronic decay channels may have on the spectrum. Indeed, it is a feature of resonances with strong $S$-wave thresholds that the states are drawn close to their strongly coupled thresholds~\cite{vanbeveren} as often discussed for the $f_0$ and $a_0$ close to ${\overline K}K$ threshold ~\cite{f0}.
Eichten, Lane and Quigg (ELQ)~\cite{ELQ} have calculated the effect of open channels
for states with hidden charm in a scheme that many find unfamiliar. In this note we want to revisit an approach related to the Dyson summation for the inverse meson propagator. This idea is not new and was considered for charmonium many years ago by Heikkil\"a {\it et al.}~\cite{tornqvist}.
What is new here is the straightforward way in which we can estimate the effects of open and nearby closed channels.

The inverse boson propagator, ${\cal P}(s)$, is shown in Fig.~1, where $s$ is the square of the momentum carried by the propagator. With $\Pi(s)$ the contribution of 
hadron loops, the complex mass function ${\cal M}(s)$ is related to this by
\bea
\nonumber
{\cal P}(s)\,\equiv\, {\cal M}^2(s)\,-\,s&=& m_0^2\,-\,s\,+\,\Pi(s)\\[1.5mm]
&=&m_0^2\,-\,s\,+\,\sum_{n=1}\,\Pi_n(s)\; ,
\eea
where $m_0$ is the mass of the bare state and the sum is over all loops (Fig.~1).  The propagator, ${\cal P}^{-1}(s)$, will then have a pole at (at least one) complex value of $s\,=\,s_R$. This position specifies  the mass and width of the physical resonance. If we denote the threshold for the $n$th channel by $s=s_n$,
then clearly only those that are open for $s \simeq {\rm{Re}}(s_R)\,>\, s_n$
contribute to the decay width of the physical hadron. However, in principle all hadronic channels contribute to its mass. Indeed, each of the infinity of closed channels contributes to the real part of $\Pi(s)$ and for a given physical mass can be thought of as redefining
the \lq\lq bare'' mass.
Since we are interested only in mass-shifts, let us subtract Eq.~(1) at some suitable point $s=s_0$ to be defined below, then
\bea
\nonumber{\cal M}^2(s)\,-\,{\cal M}^2(s_0)&=&\Pi(s)\,-\,\Pi(s_{0})\\[1.mm]
&\equiv &\sum_{n=1}\,\left[\Pi_n(s)\,-\,\Pi_n(s_0)\right]\quad .
\eea
 Since $\Pi_n(s)$ will be effectively constant for those {\it virtual} channels for which
${\rm Re}(s_R) \,<<\,s_n$, their contribution will cancel out in Eq.~(2). Consequently, the mass-shift is entirely given by the hadronic channels that are fully open or only just \lq\lq virtual''.
It is a reasonable expectation that deeply bound states, like the $J/\psi$, have masses
defined by the charmonium potential. The mass of the $J/\psi$ then essentially defines the mass scale and fixes the charm quark mass at the relevant scale.
It is thus natural to set $s_0 = M(J/\psi)^2$. In line with expectation our results change little if we use $s_0 = 4\,m_c^2\,$ instead.
For each state we take the value of ${\cal M}(s_0)$ to be that predicted by
a charmonium potential, unperturbed by hadronic channels. Of course, if the
parameters in the charmonium potential are fixed with reference to physical states for which open charm channels may contribute, we have an issue of double counting. We believe that by fixing the charmonium parameters by only deeply bound states we avoid this problem.
%
\begin{figure}[b]
\includegraphics*[width=0.95\columnwidth]{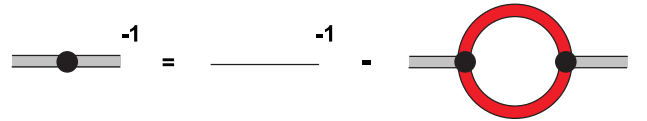}
\caption{The bare bound state propagator is dressed by hadronic
loops. The dot signifies the dressed propagator and vertices.}
\end{figure}
%

Since each $\Pi_n(s)$ is an analytic functions with a right-hand cut, we can write a Cauchy representation in subtracted form, so that
\bea
\nonumber\Delta\Pi_n(s, s_0)&\equiv&\Pi_n(s)\,-\,\Pi_n(s_0)\\[1.mm]
&=&
\frac{\left(s-s_0\right)}{\pi}\,\int_{s_n}^\infty\,ds'\,\frac{{\rm Im} \Pi_n(s')}
{\left(s'-s\right)\,\left(s'-s_0\right)}\; .
\eea
Then 
\be 
\sum_{n=1}\Delta\Pi_n(s,s_0)\,=\, {\cal M}^2(s)\,-\,{\cal M}_{charmonium}^{\,2} \;\equiv\;\Delta{\cal M}^2(s)\, .
\ee
The form of Im$\Pi_n$  for particle $P$ coupling to each channel $AB$ is taken to have a simple form, for $s \ge s_n$:
\be
{\rm Im} \Pi_n(s)\;=\;-\;g_n^{\,2}\, \left(\frac{2k}{\sqrt{s}}\right)^{2L+1}\,\exp\left(-\alpha\,k^2\right)\quad ,
\ee
where $g_n$ is the coupling of particle $P$  to channel $n$ ({\it i.e} to particles $A$ and $B$), $L$ is the orbital angular momentum between $A$ and $B$, while $k$ is the 3-momentum of $A$ and $B$ in the rest frame of $P$.
So as usual 
\be 
4k^2/s\,=\,1\,-\,2 (m_A^2+m_B^2)/s\,+\,(m_A^2-m_B^2)^2/s^2\quad .
\ee
The scale factor $\alpha$ is related to the radius of interaction, $R$, by
$\alpha=R^2/6$. This is in turn related to the size of the overlap between the $c \overline c$ and the $AB$ states. A larger value of $\alpha$ produces a smaller mass shift.  A value of $\alpha=0.4 \,\mathrm{GeV}^{-2}$ is favoured solely because it gives the most sensible results. This corresponds to $R \simeq 0.3$ fm.

For open channels, the coupling $g_n$ is simply related to the channel $n$ decay-width through Eq.~(3) with $s \simeq s_R$. For nearby closed channels we use the coupling to states with the same quantum numbers.
As a guide to the size of the effects, the calculations presented here systematically include
the channels $D{\overline D}$, $D {\overline D^*}$, $D^* {\overline D^*}$, and $D_s{\overline D_s}$.

\begin{figure}[t]
\begin{center}
\includegraphics*[width=0.95\columnwidth]{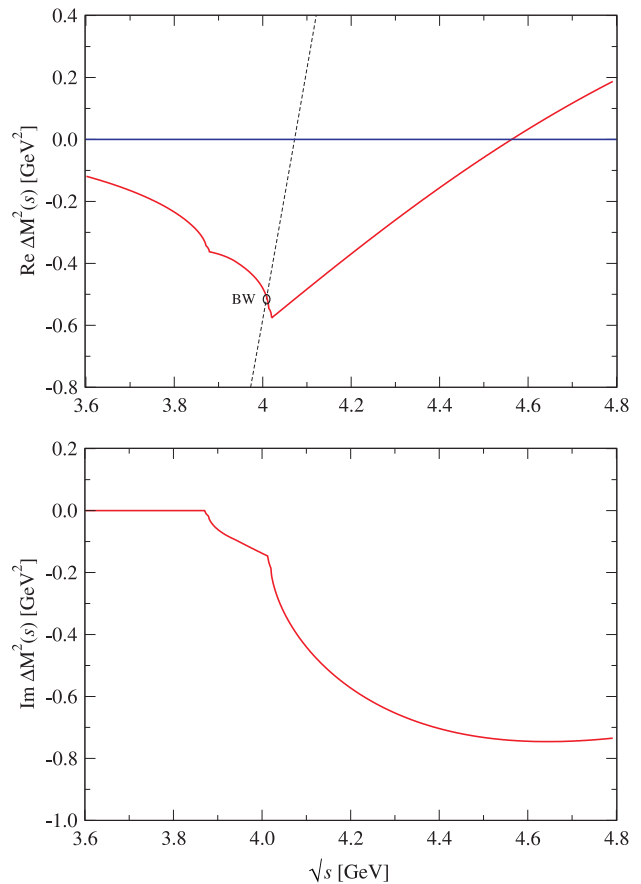}
\caption{ 
The real and imaginary parts of $\Delta{\cal M}^2(s)$ as functions of 
$E = \protect \sqrt{s}$ for the $\psi'''(3^3S_1)$ propagator. 
The dashed line shows the curve
$m_0^2-s$, where it intersects the real part of $\Delta{\cal M}^2(s)$
 labelled $BW$, defines the Breit-Wigner mass. The cusps in the real and imaginary parts occur at each of the thresholds.}
\end{center}
\vspace{-2.mm}
\end{figure}

\section{Comparison with Experiment and Other Studies}
\begin{table*}
\begin{center}
\begin{tabular}{|c c||c|c|c||c c|}
\hline
Name & State    & Experimental Mass  & Potential Mass & $\,\Gamma_{\mathrm{hadrons}}\,$&$\,\Delta m_{BW}\,$ & $\,\Delta m_{pole}\,$  \\[0.5mm]
     & $n ^{2S+1} L_J$  & MeV & MeV & MeV & MeV & MeV \\[0.3mm]
\hline
\hline
$\eta_c$ & $1^1S_0$ & 2980$\pm$1 & 2982  &--		&--               &--                 \\
$J/\psi$ & $1^3S_1$ & 3096.9     & 3090  &--		&--               &--                  \\
\hline
$\eta_c'$& $2^1S_0$ & 3638$\pm$4 & 3630  &--		&-10             & -10                \\
$\psi'$  & $2^3S_1$ & 3686.1     & 3672  &--		&-9              & -9                 \\
\hline
$h_c$    & $2^1P_1$ & 3525.9     & 3516  &--		&-2              & -2                 \\
$\chi_{c_0}$& $2^3P_0$ & 3414.8     & 3424  &--		&-9              & -9                 \\
$\chi_{c_1}$& $2^3P_1$ & 3510.7     & 3505  &--		&-16             & -16                \\
$\chi_{c_2}$& $2^3P_2$ & 3556.2     & 3556  &--		&-6              &  -6                \\
\hline
$\eta_c''$& $3^1S_0$ & 3943$\pm$6 & 4043  &	80$^a$		&-45             & -58                \\
$\psi'''$ & $3^3S_1$ & 4039$\pm$1 & 4072  &	80$\pm$10$^b$	&-36             & -41                \\
\hline
          & $3^1P_1$ &--          & 3934  &	87$^a$		&-5              & -12                \\
          & $3^3P_0$ &--          & 3852  &	30$^a$		&-70             & -70                \\
          & $3^3P_1$ &--          & 3925  &	168$^a$		&-66             & -29                \\
          & $3^3P_2$ &--          & 3972  &	80$^a$   	&-55             & -48               \\
\hline
& $3^1D_2$ &--          & 3799  &	-		&--               &--                  \\
$\psi''$& $3^3D_1$ & 3771$\pm$2 & 3785  &	23$\pm$3$^b$	&-40             &--40                \\
& $3^3D_2$ &--          & 3800  &	-		&--               &--                  \\
& $3^3D_3$ &--          & 3806  &	-		&--               &--                 \\
\hline
\end{tabular}
\end{center}
\caption{Our Results. The calculated shifts for both the Breit-Wigner and pole masses are computed from a base defined by the non-relativistic model of Barnes, Godfrey and Swanson~\protect{\cite{BGS}} listed above. Far from their masses, these predictions may be incorrect. Experimental data are from PDG~\protect{\cite{pdg}}. Only experimental errors greater than 1 MeV are quoted.\\ ${\bf ^a}$BGS~\protect{\cite{BGS}}, ${\bf ^b}$PDG~\protect{\cite{pdg}}.}
\end{table*}

We define the base from which the shift due to decay channels is to be computed  
by a non-relativistic potential model for charmonium. From the many potential modellings we choose the classic work of 
Godfrey and Isgur~\cite{GI}, more recently tabulated by Barnes, Godfrey and Swanson~(BGS)~\cite{BGS}. This is because BGS not only provide
a prediction for the eigenstates, but  include calculations using the $^3P_0$ 
model for the partial widths. It is these that fix the couplings $g_n$, which are the essential input into Eq.~(5) for computing the mass-shift from decay channels. Using these inputs  we compute the correction to the real and imaginary parts of $\Delta {\cal M}^2$ as shown in Fig.~2 for the example of the $\psi'''$. From such plots we arrive at the mass-shifts given in Table I. These are presented in two ways.
The simplest is the shift in what we call the Breit-Wigner mass, for which we only need to compute
$\Delta\Pi(s)$ at $s= {\rm Re}{\cal M}^2(s)$, Fig.~2. However, the physically relevant
quantity is the shift in the position of the pole in the complex energy plane. This requires we evaluate $\Delta\Pi(s)$ at $s\,=\,s_R\,=\,{\cal M}^2(s_R)$.
For states with small widths, of course, the Breit-Wigner and pole masses differ little.  However, for states with larger couplings, the difference is inevitably bigger. Indeed, some states get shifted below their threshold and their pole moves to the real axis. Others, however, are subject to significant changes. The largest effect is found for the $\chi_{c_1}'(3^3P_1)$ state, where a shift of $\Delta m_{BW}=-66$ MeV is reduced to just $\Delta m_{pole}=-29$ MeV.


Correcting the bare masses delivered by the potential model of Godfrey and Isgur~\cite{GI,BGS} by our calculated decay channel induced mass-shifts brings better agreement with experiment as seen from Table~I.  
 For the $\psi'''$ the downward shift by between 36 and 41 MeV is reasonable. That for the $\eta_c''$ of 45 to 58 MeV is not quite enough to bring it in line with the measured mass, which is 100 MeV below the potential model prediction.

States with common $J^{PC}$ quantum numbers have common decay channels and so
inevitably mix through these hadronic intermediate states. Two such states are
the $\eta_c'$ and $\eta_c''$. Since the ground state $\eta_c$ is deeply bound, it mixes little with these. Explicit calculation gives a shift of -0.6 MeV for the $\eta_c'$ and even less for the $\eta_c''$. Consequently, for the states listed in Tables I and II, these inter-state mixings are small and can be neglected.

\begin{table*}[t]
\begin{center}
\begin{tabular}{|c||c|c|c|c|c||c|c|}
\hline
   ~~State~~ &   ~Centroid~ & Spin Splitting &  Bare Mass & $\,\Gamma_{\mathrm{hadrons}}\,$ & $\,\Delta m_{ELQ}\,$ &   Our Mass &  Our $\Delta m$ \\[0.5mm]

          &        MeV &        MeV &        MeV &        MeV &        MeV &        MeV &        MeV \\
\hline
\hline
$2 ^1S_0$ &       3674 &      -50.1 &     3623.9 &          -- &       15.7 &     3617.0 &       -6.9 \\

$2 ^3S_1$ &       3674 &       16.7 &     3690.7 &          -- &       -5.2 &     3676.5 &      -14.2 \\
\hline
$3 ^1S_0$ &       4015 &        -66 &       3949 &         74 &        -3.1 &     3924.5 &      -24.5 \\

$3 ^3S_1$ &       4015 &         22 &       4037 &       49.8 &        1.0 &     4020.0 &      -17.0 \\
\hline
$3 ^1P_1$ &       3922 &          0 &       3922 &       59.8 &       -5.4 &     3892.0 &      -30.0 \\

$3 ^3P_0$ &       3922 &        -90 &       3832 &       61.5 &       27.9 &     3818.8 &      -13.2 \\

$3 ^3P_1$ &       3922 &         -8 &       3914 &         81 &        6.7 &     3868.9 &      -45.1 \\

$3 ^3P_2$ &       3922 &         25 &       3947 &       28.6 &       -9.6 &     3939.4 &       -7.6 \\
\hline
$3 ^1D_2$ &       3815 &          0 &       3815 &   1.7$^a$ &        4.2 &     3813.3 &       -1.7 \\

$3 ^3D_1$ &       3815 &        -40 &       3775 &  20.1$^a$ &      -39.9 &     3728.1 &      -46.9 \\

$3 ^3D_2$ &       3815 &          0 &       3815 &      0.045 &       -2.7 &     3815.0 &        0.0 \\

$3 ^3D_3$ &       3815 &         20 &       3835 &  0.86$^a$ &       19.0 &     3833.1 &       -1.9 \\
\hline
\end{tabular}
\end{center}
\caption{Comparison of the calculation and modelling by Eichten {\it et al}.
in columns 2-5, with the results from our loop calculations from their same base bare masses with their channel couplings in columns 7, 8. The couplings to individual channels are taken from the partial decay widths computed by ELQ in their Table~V of ~\cite{ELQ}. The numbers with the superscript ${\bf^a}$ are from Eichten {\it et al.}~\cite{ELQ04}.
}
\end{table*}

We first compare our calculation with that of Heikkil\"a {\it et al.}~\cite{tornqvist} of more than twenty years ago. These authors consider the spectrum of heavy quarkonia, in which the loop effects are built in from the start in the determination of the parameters of the underlying non-relativistic potential model. Meson loops then have a dramatic effect on the \lq\lq bare'' states with shifts of hundreds of MeV in mass for the lightest states. In their calculation the infinity of virtual channels (or as many of these as they choose to include) all have an effect. In contrast, in our calculation by using subtracted dispersion relations for the meson loops, the effect of the many closed channels is absorbed into the subtraction constants. Moreover, because we expect deeply bound states like the $J/\psi$ to be negligibly affected by loop corrections and well approximated by charmonium potential calculations, the subtraction constants are accurately determined.
The predicted widths by Heikkil\"a {\it et al.} are within a factor 2 of experiment for the $\psi''$ and $\psi'''$.

We now compare our results with those obtained by Eichten, Lane and Quigg~\cite{ELQ}. The first of two comparisons is directly with their results given in Table~II, columns 2-6. They ascribe part of the mass-shifts to spin splittings by suitably adjusting $\alpha_s$. 
With this the overall scale of their mass shifts assigned to decay channels is typically smaller --- compare our results in the final column of Table~I with column 6 of Table~II. For the $\psi''$, we and they find a downward shift of 40 MeV. However, the most noticeable difference is that, in our modelling, loops shift the mass downwards, whereas Eichten {\it et al.} also have appreciable upward shifts. Though the decay rates for the $^3P_0$ and $C^3$ models are qualitatively similar, there are some important differences in mass-shifts. For example, for the $\eta_c''$, the couplings are computed from decay rates at different masses.  For the BGS model, this is at a \lq\lq bare'' mass of 4043~MeV, when both ${\overline D}D^*$ and ${\overline D^*}D^*$ channels~\cite{footnote} are open. For the ELQ model, the mass is below ${\overline D^*}D^*$ threshold. Consequently, the shift in pole position for the $\eta_c''$ is $-58$~MeV in Table I, and just $-24$~MeV in Table II. For the $\eta_c'$, the same channels contribute, but both are virtual, and so we have assumed the couplings to these are as computed for the $\eta_c''$.   This results in a 10 MeV downward shift for the $\eta_c'$ using the BGS couplings, whilst only a 2 MeV shift with the ELQ~$C^3$. This would reduce the spin-spin splitting by 7 or 8 MeV. A greater reduction is required. The experimental value is 48 MeV, whilst potential models typically predict between 60 and 80 MeV.

A {\it like for like} comparison is to recompute the effect of meson loops in our calculational scheme but using the ELQ couplings and ELQ bare masses. The results are shown in Table~II. Then columns 6 and 8 can be directly compared. We see a distinctly different pattern of mass-shifts with essentially only that for the $\psi''$ being similar.
%
%
However, the calculation presented here is more straightforward to reproduce and adjust to new information on partial decay rates to be measured in the future and so may serve as a simple guide to the size of decay channel effects.

\vspace{3mm}

\noindent {\bf Acknowledgements}

\noindent DJW thanks the U.K.~Particle Physics \& Astronomy Research Council (PPARC) for a studentship. We acknowledge the partial support of the EU-RTN Programme, 
Contract No. MRTN--CT-2006-035482, \lq\lq Flavianet''.


\begin{thebibliography}{99}
\bibitem{footnote} The inclusion of charge conjugate channels is implied.
\end{thebibliography}
\end{document}